\documentclass[11pt]{article}
\usepackage{geometry}                
\usepackage{graphicx}
\usepackage{amssymb}
\usepackage{epstopdf}

\newcommand{\der}[2]{\frac{{\mathrm d} #1}{{\mathrm d} #2}}
\newcommand{\vectord}[2]{\left[\begin{array}{c}#1\\#2\\ \end{array} \right]}

\def\mlr {Marc Lachi\`eze-Rey}
\def\gr {general relativity}
\def\eg {{e.g.}}
\def\ie {{i.e.}}
\def\guill {\textquotedblleft ~}
\def\wrt  {w.r.t.}
\def\mink {Minkowski}

\def\spt {space-time}

\def\FrL{Friedmann-Lema\^ \i tre}
\def\coord {coordinate}
\def\sr {special relativity}
\def\rdot {{\dot{r}}}
\def\bydef {by definition}

\def\dS{de~Sitter}
\def\d {{\rm d}}

\title{Cosmology in the Solar System: Pioneer effect is not cosmological}
\author{\mlr}

\begin{document}
\maketitle

\abstract{
Does  the Solar System and, more generally, a gravitationally bound system follow the cosmic  expansion law    ? Is there a cosmological influence on the dynamics or optics in such systems ? 
The \gr ~theory provides an   unique  and unambiguous answer,  as a      solution of  Einstein equations with  a local source in addition to the   cosmic fluid, and obeying  the correct (cosmological) limiting conditions. This solution has no analytic expression. A Taylor development of its   metric  allows  a complete treatment of  dynamics and optics in gravitationally bound systems, up to the size of galaxy clusters,  taking into account   both  local and cosmological  effects. 
In  the solar System,  this provides an  estimation of the (non zero) cosmological influence on the Pioneer probe: it   fails to account for the \guill Pioneer effect " by about 10  orders of magnitude. We criticize   contradictory claims on this topic.
 }
 
\section{Introduction}

The Pioneer effect \cite{Anderson}  is often expressed as   an \guill anomalous acceleration " $a _{Pionneer} =c~\der{z}{t} $ in the direction of the Sun, felt by the Pioneer probe (and similar). 
The numerical  coincidence with the quantity  $c~ÊH_0^{-1} $   has led to recurrent claims of a possible cosmological origin. 
We show that (according to \gr)  this fails by several orders of magnitude. The reason is  simple :  the \guill  cosmic acceleration " of  a probe (at distance $D$)  following   the cosmic  expansion  is not $c~ÊH_0  $ but $\approx  ÊH_0^{2}~D $. For the Pioneer probe, this amounts to   about 10 orders of magnitude below  $a _{Pionneer}$.

Although this conclusion can be expected from order of magnitude arguments (see section \ref{trois}),  it is the   rigorous consequence of the    more general analysis  performed in \cite{mizo}. We gave there   the   solution of  \gr ~which   describes  a gravitationally bound  system   (the Sun, a galaxy, a galaxy cluster, ...) in the cosmological environment:  this is the solution of   the Einstein equations with the local (like the Sun) and global (the homogeneous isotropic   cosmic fluid)  gravitational sources; the  limiting conditions   are not those of a flat (\mink) \spt, but  tend  asymptotically toward    the desired cosmological model, like the present favorite $\Lambda$-CDM.  This system has no analytical solution in general (excepted  the Schwarzschild - de Sitter solution for   a pure de Sitter cosmological model, and with  spherical symmetry). However, \cite{mizo}  provided  a Taylor  development  of the resulting  metric     in small parameters:  the local (Newtonian) gravitational potential $\phi$, and a dimensionless parameter $r~H_0/ c  $,   expressing the characteristic dimension of the system in  Hubble length  units.
This approximation replaces the  unknown exact   \spt ~solution by an  \emph{osculating \spt} ~(\cite{mizo}, see also \cite{Bel}), which is shown to describe   perfectly  the Solar system,   galaxies, and galaxy clusters.

This enlightens  the question of the influence of cosmology in these    systems (see also   \cite{CarreraF}). 
In particular,   an inertial probe  in radial motion in the  Solar System feels  a \guill cosmic acceleration " (which is defined below covariantly)    $q_0~ÊH_0^{2}~D $. Its magnitude is about 10   orders of magnitude below   the Pioneer effect. In the last section, we analyze some works  by    \cite{Oliveira}, \cite{Fahr} and  \cite{Carrera} about cosmological influence on the Pioneer probe.

 \section{The osculating metric}

This section gives a short account of the results of \cite{mizo}, in the framework of \gr.
  
 The chrono-geometry of  a gravitationally bound  system   is   correctly described  
by the   solution of  the Einstein equations with the local (\eg, the Sun) and cosmic   sources, which tends  asymptotically toward  the  \FrL ~model, with the  cosmic expansion law involving the  measured values of the cosmic parameters ($H_0, ~\Omega_M,~\Omega_\Lambda$).
No analytic solution exists in general but we  define  two small parameters: \begin{itemize}
  \item  the local gravitational potential $\phi$. A first   (Newtonian)  order analysis is sufficient if we are not interested in strong field (post-Newtonian) effects;
  \item   and a    small dimensionaless parameter $r~H_0  $ (we chose now units where $c=1$), characterizing the influence of cosmology. Estimations show that this parameter remains very small: from about $10^{-13}$ in the solar system to $10^{-3}Ê$ in a galaxy cluster.
\end{itemize}

For the case of  a central massive  source,   \cite{mizo} have  obtained the  Taylor   expansion of the metric (their equation 22) at first order in 
 $\phi$  and second cosmological order:
\begin{equation}
\label{dsU2} g\approx    \left( 1-2\,{\frac {M}{r}}+q_0~(H_0 r)^{2}  \right) {{\it dt}}^{2}-{{ \it dr}}^{2} \left( 1+2\,{\frac {M}{r}}+\Omega _0\,(H_0 r)^{2}\right)  
 -{r}^{2}{{\it d{\omega}}}^{2}.\end{equation}
Here $H_0$ is the present Hubble constant (not the time varying Hubble parameter) and $q_0$ the present deceleration parameter.  This  osculating  metric  \cite{mizo}  at the position of the observer solves  the  Einstein equations  at the desired order. It is   written in a convenient  static  \coord  ~system. Putting $ \phi=0$ gives the pure \FrL ~model in its  static form, see \cite {mizo}. Putting  $H_0  =0$ gives the usual    Newtonian solution, without cosmological effects. The calculations  assume  spherical symmetry of the source.  When the cosmology is pure \dS, this solution is exact.

{\bf The Pioneer effect is not cosmological}

The motion of an inertial   probe like Pioneer is described by the radial geodesic equation derived from the  metric (\ref{dsU2}). So is  the null radial geodesic equation corresponding to the light-rays from the probe to the observer.   This provides  the redshift $z(\tau)$ of the probe measured by the observer,  as a function of his proper time $\tau$, and its derivative $\dot{z}\equiv \der{z}{\tau}$. All these quantities are covariantly defined. 

A   comparison   with the pure   Newtonian treatment (\ie, which neglects   cosmology)   must   involve    frame-invariant   quantities only.  This is the case for   $z$, $\tau$ and  $\dot{z}$.  This latter observable  quantity    is called \guill acceleration ", and the difference with  the pure Newtonian estimation, which measures the effect of cosmology,  is called   $a_{cosmic}$. The calculations  \cite{mizo}  give   $ a_{cosmic} =q_0~ÊH_0^2~r$, with $q_0$ the usual deceleration parameter, ten orders of magnitude below $a_{Pioneer}$. This  expression is exact even at the second cosmological  order.  The motions of  both the source and  the electromagnetic signal (in the optical approximation) are treated as relativistic, in the \spt ~curved by the cosmology and by the local potential. 
  All calculations are covariant.  

 \subsection{Orbital motion: a modified Kepler law}
 
To complete the discussion of local cosmological effects, we consider the case of orbital  motion, assumed purely   circular:  $V^r=V^ \varphi=0$. From   \cite{mizo}, we may explicit the geodesics equations in the metric above:
 $$\left( 1-2\,{\frac {M}{r}}+q_0~(H_0 r)^{2}  \right)~ÊV^t~ÊV^t-r^2~ÊV^\theta~ÊV^\theta=1,$$
 $$\left(    {\frac {M}{r ^2}}+Êq_0~ H_0 ^{2}~ rÊ   \right)~ÊV^t~ÊV^t - r ~ÊV^\theta~ÊV^\theta=0.$$
They lead  to 
$$V^t\approx 1+\frac{3M}{2r},$$
$$(r~V^\theta)^2\approx \frac{M}{r}+q_0~(H_0r)^2.$$
These quantities are the components of a vector. They  are not  directly observables, nor covariant.
However, we may characterize an orbit by its proper period $T$ and proper circumference $C=2\pi ~r$.

The proper period is the integral of proper time along a closed orbit.
$$T = \int _{orbit}Ê\d s=\int _{0}^{2\pi}Ê\d \theta \der{s}{\theta}=\frac{2\pi}{V^\theta}
=\frac{2\pi ~r}{ \sqrt{ \frac{M}{r}+q_0~(H_0r)^2}}. $$
  
This  may be expressed under the form of   a modification of the   Kepler law
\begin{equation}
\label{Kepler}T^2~Ê(1+\epsilon_{cosmo})=\frac{C^3}{2\pi ~ÊM}; ~Ê
\epsilon_{cosmo}\equiv \frac{q_0~ÊH_0^2~ÊT^2}{(2\pi)^2}.
\end{equation}
 
This accounts for the cosmological effect on  a circular orbital motion in a gravitationnally bound system  (see also the discussions by \cite{CarreraF} and references therein), at Newtonian order, and second cosmological order:  the orbit  is not expanding, nor shrinking (at first order). Cosmology slightly  modifies its characteristics   \wrt ~the pure Newtonian case. We have expressed this modification as a modification (\ref{Kepler})  of the Kepler law. 
This effect is far from being  measurable in the Solar system.

 \subsection{Cosmological effects  in gravitationally bounded systems}

The osculating metric  describes  any  gravitationally bound system, like   galaxies, galaxy clusters...,    clearly  accounting for the   cosmological influence.
Although calculations are performed with spherical symmetry, there is no problem (beside technical) to extend them to any distribution of sources.  The case  of  an object which is bound by non gravitational interactions, like a material rod, an hydrogen atom..., remains open. 
We summarise our main results

\begin{itemize}
  \item 
In such a bound    system,   an inertial  object   suffers basically the same cosmological effect    that it would suffer in the absence of local gravitational sources. This effect may be described by a \guill cosmic acceleration " which adds to the local terms. Note that this cosmic acceleration would be zero for a non accelerated  expansion  ($q_0=0$). 
  \item 
For application to the Solar systems, galaxies, galaxy clusters,   a  first order expansion in the cosmological parameter is largely sufficient (our results are valid up to second order).

   \item 
For a  probe in   radial [inertial] motion, this acceleration is manifest   as an additional term in  the derivative of the redshift \wrt ~the observer's proper  time. This is qualitatively of the same nature of the Pioneer effect, although many orders of magnitude below (see also the following section and \cite{Carrera} for a kinematical cosmic contribution).
  \item 
  For a probe 
in  orbital [inertial] motion (\ie, a planet), the cosmic acceleration adds to the local terms and slightly modifies the  orbit, whose characteristics remain however  constant (at first order): it  does not expand or shrink (up to second order) under the effect of cosmology. The   difference with  the pure  Newtonian case may  be expressed   as  a modification (\ref{Kepler}) of the Kepler law.
  \item 
In the Solar system,  the  cosmic acceleration is   far from being measurable. It fails to account for  the Pioneer effect.  
\end{itemize}

The  weak   influence of cosmological effects, measured by  $q_0~ÊH_0^2~r$,    increases with the size of the system. In external regions of galaxies, or galaxy clusters, 
  precision analyses (like estimations of dark matter, gravitational lensing.... ) would require  to take it  into account. 
Since their signature differs from that of the local effects, it may be hoped that future observations will reach the precision   allowing  to distinguish them from the  local effects, so   providing new kinds of  cosmological tests.  
 
\section{Different analyses of the Pioneer effect }\label{trois}

For the moment, the approximate coincidence between $a_{Pioneer}$ and $c~ÊH_0$ has no explanation. We have shown that it cannot be explained by relativistic cosmology. A   modification  of   our gravitational theory may possibly  provide an explanation for the Pioneer effect   (\cite{Reynaud},\cite{Iorio} and references therein)  but the answer would not be cosmological.

To complete this short note, we analyse some  recent discussions about    cosmic effects  for the Pioneer probe. One is in the context of special relativity  \cite{Oliveira}  ;  another  in that of   a pure expanding cosmology \cite{Fahr}. Both neglect  the  local gravitational sources. 
We show that  even in such  simplified approaches,   the correct treatment   leads to the  conclusion  that cosmology fails by several orders of magnitude to account for the Pioneer effect.

\subsection{A special relativistic approach}

In the framework  of \sr,  a particle following an expansion law  cannot be inertial, but   is submitted to a \guill  cosmic force ". The latter   imprints an acceleration $a_{cosmic}$  responsible for the Hubble law, that \cite{Oliveira}  compare    to $a_{Pioneer}$. 

The authors  start  from an application of the redshift law to the de Broglie length of a massive particle. Unfortunately, this appears to be   incompatible with   \sr.   The special relativistic formula for  the redshift of  a particle,  $z = v/c = \rdot /c$,   implies for the de Broglie  length $ \lambda = h/p=h/mv\propto \frac{1}{ z}$: it   does not follow the expansion law. Moreover, it is impossible to imagine any modification   justifying   formula (9) of \cite{Oliveira}, since the latter  would   imply  $z \mapsto \infty $ for $v=0$, instead of the correct formula $z=1$. There is no way to reconcile an expansion law for the de Broglie length with \sr. 
 
 In fact, for     a particle following an  Hubble law, the redshift $z=v=\der{D}{t}=H~D$.
Derivation gives
$$\der{z}{t}= a_{cosmic}= \der{H}{t}~ÊD+H~\der{D}{t}=-q_0~ÊH^2~ÊD,$$
in accordance with the  result  above. 

Note  the correct  cosmological evolution  of  the de Broglie length length  of  a \emph{non massive} particle ($p= h \nu/c$),   $$\lambda =\frac{c}{  \nu}= \frac{c  }{   \nu _0 + \delta \nu }=(1+z)~Ê\lambda _0 .$$
 
 \subsection{Pure cosmic expansion approach}

A different approach is adopted in \cite{Fahr}, who also neglect the local potential. 
The   relativistic approach of their section (2), involving  geodesics,     gives the correct order of magnitude. Their section (3), however,   invokes a  possible cosmic explanation for  the Pioneer acceleration. We show that it is based on a misinterpretation of the cosmological equation.

For an  observer at cosmic  time $t_{obs}$, the equation (25) of  \cite{Fahr}     must be written 
\begin{equation}
\label{zz}
1+z(t_{obs})=\frac{\lambda_1}{\lambda_0}=\frac{a(t_{obs})}{a[t_{source}(t_{obs})]},
\end{equation}
where $t_{source}(t_{obs})$ is the    time of emission, by the source, of the light-ray reaching the observer at $t_{obs}$. It is  solution of the null radial geodesic equation: 
\begin{equation}
\label{integral}
\int _{t_{source}}^{t_{obs}}~\frac{ \d t'}{a(t')}=r,
\end{equation} with $r$   the constant \coord ~of the comoving source. 
Derivation of the latter  gives \begin{equation}
\label{derr}
 \der{t_{source}}{(t_{obs})}=\frac{a( t_{source})  }{ a(t_{obs})}.
\end{equation}

 On the other hand, the Taylor  development  of (\ref{zz}) near $t_{obs}$ gives
\begin{equation}
\label{zzex}
 z(t_{obs})=-H(t_{obs})~(t_{source}-t_{obs})
+H(t_{obs})^2~(t_{source}-t_{obs})^2~(1+q_0/2).
\end{equation}

Derivation gives
\begin{equation}
\der{z(t_{obs})}{t_{obs}}= 
-\der{H(t_{obs})}{t_{obs}}~(t_{source}-t_{obs})
-H(t_{obs})~(\der{t_{source}}{t_{obs}}-1)\end{equation}
$$+2 ~ÊH(t_{obs})~\der{H(t_{obs})}{t_{obs}}~(t_{source}-t_{obs})^2~Ê(1+q_0/2)
+2 ~ÊH(t_{obs})^2~(t_{source}-t_{obs})~(\der{t_{source}}{t_{obs}}-1)~Ê(1+q_0/2),
$$ where the two last terms are at third order. Thus, after using (\ref{derr}) and (\ref{zz}), we obtain again
(up to  second order) $$\der{z(t_{obs})}{t_{obs}}= -q_0~r~a(t_{obs})~ÊH(t_{obs})^2.$$

This result could be derived  more quickly     :  neglecting local gravitation,   a comoving galaxy is at a proper distance $D(t)=a(t) ~Êr$ from the observer, with    $r$ the    comoving \coord ~Êof the source,    constant \bydef. 
The Hubble law follows:  $V = \dot {D}=H ~Êa(t)~  r=H ~ÊD(t)$. 
Derivation gives $\dot{V}=\dot{H}~ÊD+H~Ê\dot{D}$, which leads to  $a_{cosmic}$ as above.

Thus, at the required order,  pure cosmological calculations (\ie, without local effects) are sufficient to provide   the correct value of the cosmic acceleration. In the same  context,     
\cite{Carrera} also exhibited an additional contribution of cosmological origin.
  
\subsection{A cosmologic  kinematical effect}

In addition to the \guill cosmological acceleration " considered above,  cosmology also imprints a kinematical signature of the time-derivative of the redshift, as it  was  recently shown   by     \cite{Carrera}.

In the pure  cosmological context (neglecting the local gravitation), 
described by the usual RW metric, the comoving   observer  has velocity    $u=\vectord{1}{0}$.
An  arbitrary radially moving source has  velocity    
$v =\vectord{\gamma (t)}{\sigma (t) /a(t)}$, with
 $\gamma (t)^2 -\sigma ^2(t)=1$, and   $a(t) $ the usual cosmic scale factor.
The comoving   observer  observes the source with a  redshift  $1+z=\frac{\delta t_{obs}}{\delta t_{source}} =\frac{a(t_{obs})}{a(t_{source})}~Ê(\gamma+\sigma)$.

The case of   signal reflection has been studied in more details by \cite{Carrera}: 
a signal  emitted by the comoving observer at $t_0$, reflected by a moving mirror  at $t_1$ (with velocity $v$ as above)  is observed by the   same  comoving observer at $t_2$ with a redshift
\begin{equation}
\label{redshift} 1+z_{20}=\frac{\delta t_2}{\delta t_0} 
=\frac{a(t_2)}{a(t_0)}~Ê(\gamma+\sigma)^2.
\end{equation}
 This is both   the    formula    (17) and the    formula (22)  of \cite{Carrera} (they are identical for   radial motion).

In the inertial case (relevant for  the Pioneer probe),  $\gamma$ and $\sigma$   remain constant. 
   It is easy to calculate  the derivative of the redshift (\ref{redshift})  \wrt ~$t_2$, which is the proper time of the observer.
Using the usual Taylor developments of the cosmic quantities, one obtains  (at lowest order) 
$$\der{ z }{t_2}\approx  H_2^2~(1+q ) ~(t_0-t_2)+z   ~H_2 , $$ with $H_2$ and $q$    the Hubble constant and the deceleration parameter at observer's position.

The  first term in the RHS is the cosmic acceleration   discussed above.
The  second term represents  the  kinematical contribution
pointed by \cite{Carrera}.   As discussed by these authors, it  is also  negligible  in the situation  of the   Pioneer probe.

\end{document}